\documentclass[11pt]{article}

\usepackage{booktabs}
\usepackage[english]{babel}
\usepackage{amsmath,amssymb,amsbsy,amstext, amsthm, simplewick}
\usepackage{hyperref}
\usepackage{graphicx}
\usepackage{amsfonts}
\usepackage{amssymb}
\usepackage[small]{caption}
\usepackage{upgreek}

\usepackage{colortbl}
\definecolor{lightgreen}{cmyk}{0.2, 0, 0.2, 0.2}
\definecolor{lightgray}{cmyk}{0.1,0.2,0,0.1}
\definecolor{lightgray2}{cmyk}{0.1,0.1,0,0.1}

\makeatletter
\newlength{\apb@width}
\newcommand{\autoparbox}[2][c]{\settowidth{\apb@width}{#2}\parbox[#1]{\apb@width}{#2}}

\makeatother

\numberwithin{equation}{section}

\def\beq{\begin{equation}}
\def\eeq{\end{equation}}

\def\bea{\begin{eqnarray}}
\def\eea{\end{eqnarray}}

\def\beq{\begin{equation}}
\def\eeq{\end{equation}}
\def\bea{\begin{eqnarray}}
\def\eea{\end{eqnarray}}

\def\fNL{f_{\rm NL}}
\def\Mp{M_{\rm pl}}

\def\fNLb{\bar{f}_{\rm NL}}

\def\k{{\bf k}}

\DeclareRobustCommand{\SkipTocEntry}[4]{}

\begin{document}

\begin{titlepage}

\setcounter{page}{1} \baselineskip=15.5pt \thispagestyle{empty}

\bigskip\

\vspace{2cm}
\begin{center}
{\fontsize{18}{32}\selectfont  \bf Testing Split Supersymmetry with Inflation}
\end{center}

\vspace{0.5cm}
\begin{center}
{\fontsize{14}{30}\selectfont   Nathaniel Craig$^{\spadesuit}$ and Daniel Green$^{\blacklozenge,\clubsuit}$}
\end{center}

\vspace{0.2cm}

\begin{center}
\vskip 8pt
\textsl{$^\spadesuit$ Department of Physics and Astronomy, Rutgers University, Piscataway, NJ 08854, USA}
\vskip 7pt
\textsl{$^ \blacklozenge$ Stanford Institute for Theoretical Physics, Stanford University, Stanford, CA 94305, USA}
\vskip 7pt
\textsl{$^\clubsuit$ Kavli Institute for Particle Astrophysics and Cosmology, Stanford, CA 94305, USA}
\end{center}

\vspace{1.2cm}
\hrule \vspace{0.3cm}
{ \noindent {\sffamily \bfseries Abstract} \\[0.1cm]
Split supersymmetry (SUSY) -- in which SUSY is relevant to our universe but largely inaccessible at current accelerators -- has become increasingly plausible given the absence of new physics at the LHC, the success of gauge coupling unification, and the observed Higgs mass.  Indirect probes of split SUSY such as electric dipole moments (EDMs) and flavor violation offer hope for further evidence but are ultimately limited in their reach.  Inflation offers an alternate window into SUSY through the direct production of superpartners during inflation.  These particles are capable of leaving imprints in future cosmological probes of primordial non-gaussanity.  Given the recent observations of BICEP2, the scale of inflation is likely high enough to probe the full range of split SUSY scenarios and therefore offers a unique advantage over low energy probes.  The key observable for future experiments is equilateral non-gaussianity, which will be probed by both  cosmic microwave background (CMB) and large scale structure (LSS) surveys.  In the event of a detection, we forecast our ability to find evidence for superpartners through the scaling behavior in the squeezed limit of the bispectrum.
\noindent}
\vspace{0.3cm}
\hrule

\vspace{0.6cm}

\end{titlepage}

\newpage 

\section{Introduction}

The Standard Model has been completed by the discovery of an apparently elementary Higgs boson at the LHC. On one hand, the absence of evidence for additional degrees of freedom at the LHC challenges many proposals for new weak-scale physics beyond the Standard Model. On the other hand, the recent discovery of primordial tensor modes in the CMB by BICEP2 \cite{bicep2} points to the existence of new physics at a scale that suggestively coincides with apparent  gauge coupling unification in supersymmetric extensions of the Standard Model \cite{Dimopoulos:1981yj}. That the scale indicated by cosmological observations coincides with the scale indicated by low-energy observations is extremely suggestive. In this paper we pursue the idea that cosmology may provide even more concrete evidence for the existence of supersymmetry (SUSY) well above the weak scale.\footnote{In some string models, having $m_{\frac{3}{2}} < H$ causes problems for moduli stabilization \cite{Kallosh:2004yh}, which some authors take as evidence against low scale SUSY.  We will ignore such concerns here. }

Cosmological inflation \cite{inflation} offers a novel opportunity to search for SUSY in the universe.  The discovery of primordial tensor modes in the CMB by BICEP2 \cite{bicep2} strongly supports the idea that inflation occurred at very high energies.  For the reported central value of $r= 0.2^{+0.07}_{-0.05}$, the inflationary Hubble scale is given by $H \sim 1.1 \times10^{14}$ GeV.  Since any field with mass less than the inflationary Hubble scale can be produced during inflation, cosmological observables are sensitive to particles produced at these incredible energies.  
 
Although the potential reach in energy of inflation is well-known, it has been less appreciated in the particle physics community that cosmological observables can directly test the presence of additional particles and interactions at these scales (see \cite{Abazajian:2013vfg} for a recent review).  One crucial observation is the single-field consistency condition \cite{Maldacena:2002vr,Creminelli:2004yq}, which states that if inflation is described by a single degree of freedom then the bispectrum of the  scalar curvature perturbation, $\zeta$, satisfies
\beq\label{equ:singlecc}
\lim_{\k_3 \to 0} \langle \zeta_{\k_1} \zeta_{\k_2} \zeta_{\k_3} \rangle' \to  P_\zeta(k_1) P_\zeta (k_3) \left[ (n_s -1) + {\cal O}(k_3^2) \right] \ .
\eeq
Deviations from the consistency condition offer a relatively clean method for detecting additional fields present during inflation.  The most commonly studied deviation is the case of local non-Gaussanity, where $(n_s - 1) \to \fNL^{\rm local}$, which is most easily produced by additional {\it massless} scalars.  On the other hand, {\it massive} scalars with $0 < m \leq \tfrac{3}{2} H$ give rise to a bispectrum with soft limit \cite{Chen:2009zp}
\beq\label{equ:multi}
\lim_{\k_3 \to 0} \langle \zeta_{\k_1} \zeta_{\k_2} \zeta_{\k_3} \rangle' \to P_\zeta(k_1) P_\zeta (k_3) \left[ (n_s -1) + {\cal O}(k_3^{\alpha}) \right] \ ,
\eeq
where $\alpha \equiv \tfrac{3}{2} - \sqrt{\tfrac{9}{4} - \frac{m^2}{H^2}}$.  Measuring $\alpha < 2$ both tells us that there is an extra degree of freedom and indicates its mass\footnote{Strictly speaking, weakly coupled massive particles only produce $\alpha \leq \tfrac{3}{2}$.  Taking $m > \tfrac{3}{2} H$ does not extend this limit, as these massive fields can be integrated out, up to exponentially suppressed contributions \cite{Chen:2012ge}.  There is no obstacle to producing the full range $0 \leq \alpha< 2$ with additional fields, as was demonstrated concretely in \cite{Green:2013rd}.  } during inflation.

The above phenomenon provides a novel search technique for supersymmetry at high scales \cite{Baumann:2011nk}. Although the non-observation of superpartners at the LHC is beginning to challenge scenarios of weak-scale supersymmetry, there remains strong motivation for so-called {\it split supersymmetry} scenarios where most or all superpartners lie outside the reach of the LHC  \cite{split, minisplit}. If this is the course Nature has chosen, verifying the existence of supersymmetry at high scales requires new experimental probes. Provided that more direct sources of SUSY breaking scale are below the inflationary Hubble parameter, then the dominant source of SUSY breaking during inflation is set by the curvature, namely $H$.  As a result, we expect to find additional scalar particles with masses set by $H$ that naturally produce signatures at a detectable level  \cite{Baumann:2011nk, Assassi:2013gxa}.  A detection of $\alpha \sim 1$ would then provide tantalizing evidence that SUSY is relevant to our universe, even if it is never probed directly at the LHC (see also \cite{Iliesiu:2013rqa} for a different approach).

In this paper, we will explore the capability of cosmological observables to shed light on SUSY at high scales.  In section \ref{sec:susy} we review the low-energy evidence in support of supersymmetry at high scales, including the success of  precision gauge coupling unification and the observed Higgs mass. In particular, the observed Higgs mass provides a suggestive upper bound on the present scale of SUSY breaking. In section \ref{sec:reach}, we then discuss the reach of cosmological observations in terms of the scale of SUSY breaking.  We will discuss the assumptions that go into the predicted signals and how these compare with existing indirect probes.  In section \ref{sec:forecast}, we will forecast our ability to detect $0 < \alpha < 2$ in an ideal 3d experiment, with an eye towards large scale structure surveys.  In section \ref{sec:fakes}, we present possible alternative explanations of such a signal and how one could try to distinguish them.  We conclude in section \ref{sec:outlook} with a discussion of the prospects for observation. Although the BICEP2 measurement of primordial tensor modes provides strong motivation, our study remains relevant irrespective of future changes in the central value of $r$.

\section{Split supersymmetry and its experimental probes} \label{sec:susy}

The apparent unification of Standard Model gauge couplings under extrapolation to higher energies has long been a suggestive indication of new physics many orders of magnitude above the weak scale \cite{Dimopoulos:1981yj}. Although gauge coupling unification in the context of the Standard Model alone is badly disfavored by both precision measurements from LEP and SLC and by the non-observation of proton decay, it is highly successful in supersymmetric extensions of the Standard Model. The additional matter content dictated by supersymmetry -- particularly electroweak doublet fermions -- places the supersymmetric prediction for gauge coupling unification in reasonable agreement with precision data and lowers the rate for proton decay consistent with current limits. It also picks out a particular scale for gauge coupling unification, $M_{\rm GUT} \sim 2 \times 10^{16}$ GeV, which is suggestively close to the scale of the inflationary potential favored by the observation of primordial tensor modes. In the conventional paradigm of weak-scale supersymmetry, supersymmetric unification can align within $\sim 3 \sigma$ of current low-energy data and may be reconciled with a modest $3-4\%$ threshold correction at the unification scale \cite{Beringer:1900zz}. However, the non-observation of superpartners at the LHC is beginning to put this paradigm under stress. 

\begin{figure}[t]
\begin{center}
\includegraphics[width=.48 \textwidth]{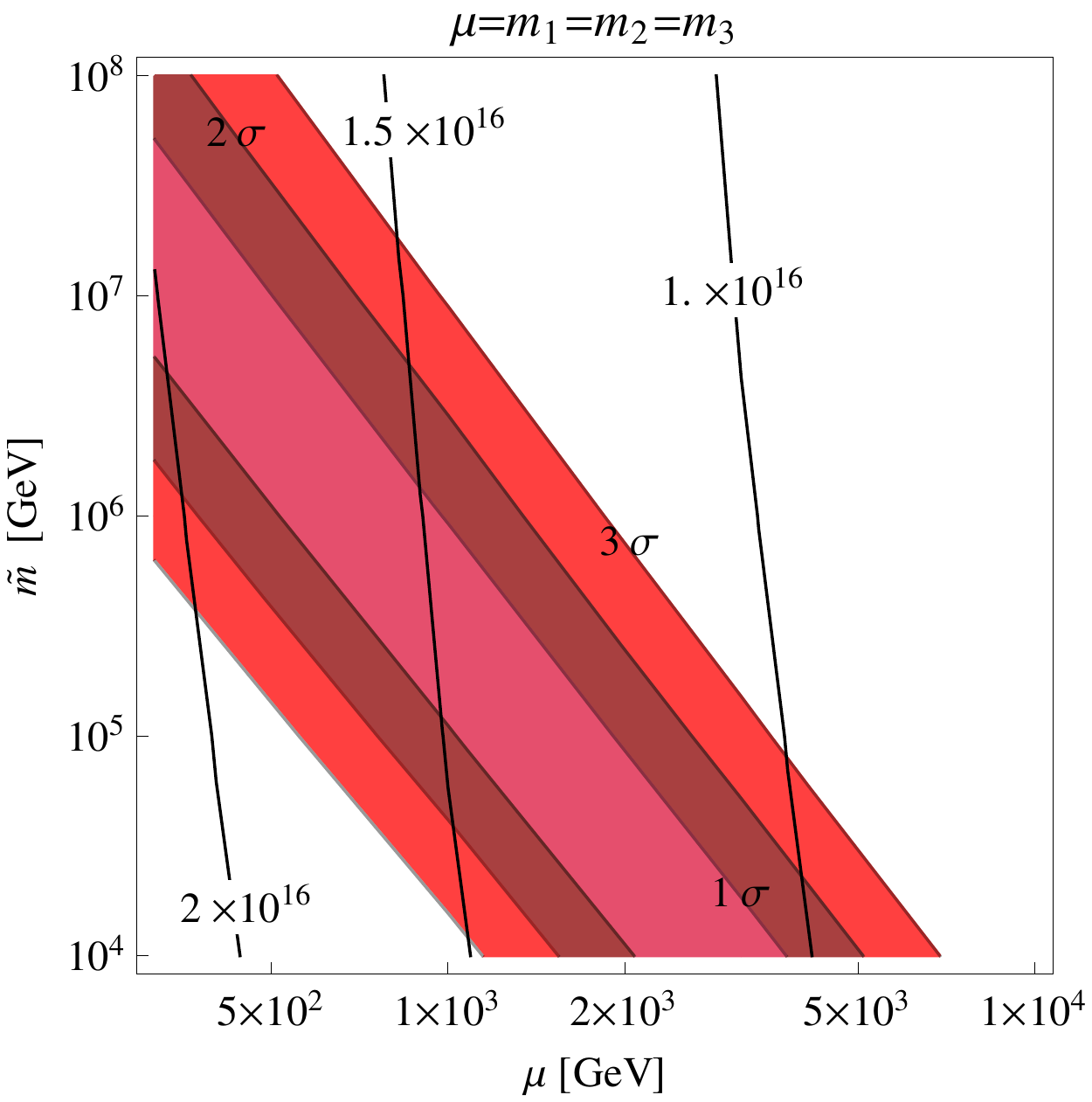}
\includegraphics[width=.48 \textwidth]{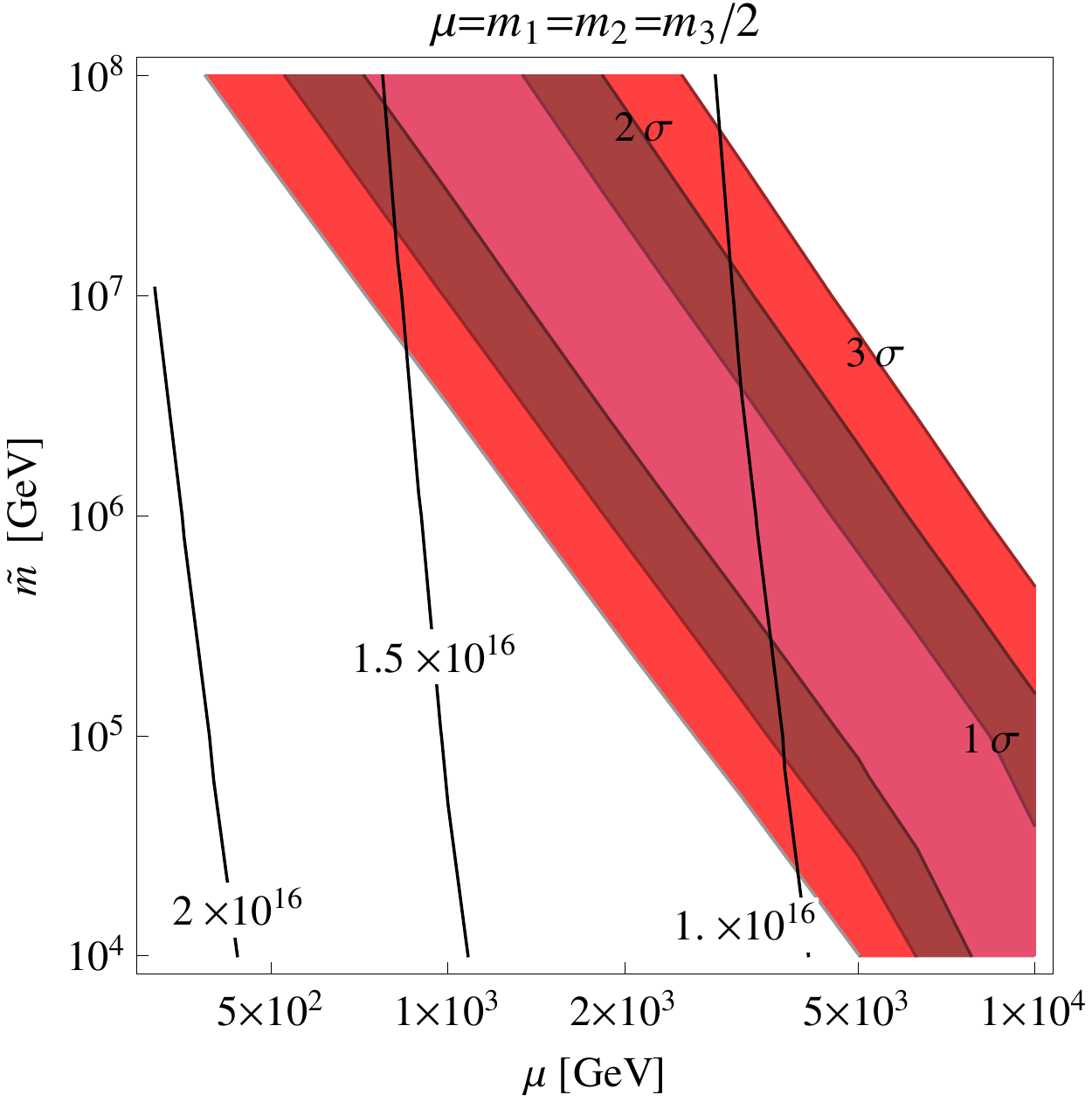}
\caption{ \label{fig:gut} \small\it Unification prediction in split supersymmetry as a function of common fermionic ($\mu$) and scalar ($\tilde m$) superpartner masses using two-loop running \cite{split}, neglecting weak-scale and unification-scale  thresholds. In the left panel we take a common physical mass scale for all fermionic superpartners, while in the right panel we take the gluino to be twice as heavy as the other fermions. The diagonal bands represent the $1\sigma, 2\sigma, 3\sigma$ constraints corresponding to $\alpha_3(M_Z)=0.1184\pm0.0007$, where we have taken the experimental inputs $\alpha_{em}^{-1}(M_Z) = 127.916$ and $\sin^2 \theta_W(M_Z)=0.23116$. The solid black contours indicate the one-loop unification scale in units of GeV.}
\end{center}
\end{figure}

Surprisingly, the success of supersymmetric gauge coupling unification is {\it improved} in split supersymmetry \cite{split}, a scenario where scalar superparters lie well above the weak scale while fermionic superpartners are further protected by an $R$-symmetry and remain light. In this scenario supersymmetry no longer accounts for the entirety of the hierarchy between the weak scale and the Planck scale, although it still protects the weak scale against radiative corrections over many decades in energy.  As the scalars are made heavy, the supersymmetric prediction for gauge coupling unification aligns perfectly with low-energy data without relying upon additional threshold corrections \cite{split}. Fermionic superpartners are favored to remain light, since the contributions to running couplings from the light superpartners of the Higgs boson, the higgsinos, are largely responsible for successful unification. The precision of the unification prediction is illustrated in figure \ref{fig:gut}, which makes apparent that unification prefers light fermionic superpartners, although the preferred mass range for both scalars and fermions depends on the details of the fermionic spectrum. The inferred unification scale depends primarily on the higgsino mass, varying weakly from $M_{\rm GUT} \sim 5 \times 10^{15} - 2 \times 10^{16}$ GeV as the higgsinos vary from $\mu \sim 10^2 - 10^4$ GeV.

The case for some form of split supersymmetry is bolstered by the observation of a Standard Model-like Higgs of mass $m_h \sim 126$ GeV. In minimal supersymmetric extensions of the Standard Model, the Higgs quartic is fixed by supersymmetry and radiative corrections due to supersymmetry breaking. In the case of split supersymmetry with light gauginos and heavy scalars, the observed Higgs mass is consistent with scalar superpartners in the range $\tilde m \sim 10^4 - 10^8$ GeV \cite{Giudice:2011cg}. This favors the scenario of ``mini-split'' supersymmetry \cite{minisplit}, in which scalars lie within six orders of magnitude of the weak scale -- a subset of the possible range available in the original incarnation of split supersymmetry.

The bound on scalar superpartners in mini-split supersymmetry is fairly robust. Extensions of the MSSM that introduce additional quartic couplings typically {\it lower} the upper bound on $\tilde m$ by increasing the tree-level prediction for the Higgs mass. It is possible to raise the bound on $\tilde m$ if $A$-terms are large enough to induce negative threshold corrections to the quartic coupling, but this typically leads to prohibitive charge- and color-breaking minima well before the mass bound is substantially weakened. Alternately, if both scalar and gaugino masses are well above the weak scale (so-called ``heavy supersymmetry'', distinct from split supersymmetry), the running of the Higgs quartic changes such that the mass bound is relaxed to $\tilde m \lesssim 10^{13}$ GeV  \cite{Giudice:2011cg}, but at the cost of sacrificing precision gauge coupling unification if fermionic superpartners are heavier than $\sim 10^6$ GeV \cite{minisplit}. 

In the framework of split supersymmetry, the mass bound on scalars can be translated into an upper bound on the scale of supersymmetry breaking in the present vacuum. There are generic Planck-scale contributions to scalar masses of order $\delta \tilde m_{\rm grav} \sim F / (\sqrt{3} M_{\rm pl}),$ which implies $\sqrt{F} \lesssim 2 \times 10^{13}$ GeV. Low-scale mediation mechanisms such as gauge mediation entail even smaller values of $\sqrt{F}$. Although these contributions may be sequestered away, there remain anomaly-mediated contributions of order $\delta \tilde m_{\rm amsb} \sim 10^{-2} \delta \tilde m_{\rm grav}$, which implies at the very least  $\sqrt{F} \lesssim 2 \times 10^{14}$ GeV. Thus the observed Higgs mass indicates that the scale of supersymmetry breaking in the present vacuum is, at its largest, two orders of magnitude below the scale of gauge coupling unification, assuming the boundary conditions for the Higgs quartic are set by split supersymmetry.

Finally, split supersymmetry may provide a viable dark matter candidate if neutral gauge fermions are sufficiently light and $R$-parity is conserved, though the dark matter candidate is subject to constraints from direct \cite{Cheung:2012qy} and indirect  \cite{Cohen:2013ama} searches. This is particularly attractive if the QCD axion with GUT-scale axion decay constant is no longer an effortlessly viable candidate for the majority of dark matter, as suggested by the combination of primordial tensor modes and isocurvature constraints \cite{Fox:2004kb}.

While these indications are suggestive of split supersymmetry, they are not decisive. Precision unification, dark matter, and a viable Higgs mass prediction can all be achieved in non-supersymmetric extensions of the Standard Model, particularly if electroweak naturalness is no longer a strong guide. Split supersymmetry may be probed directly if fermionic superpartners such as the gluino are within kinematic reach of the LHC, but this is far from guaranteed; gluinos may be kinematically inaccessible at the LHC without imperiling precision gauge coupling unification. Thus in order to determine whether supersymmetry is present at higher energies, indirect probes take on crucial significance.

At present the best indirect probes of split supersymmetry are precision observables such as electric dipole moments (EDMs) and flavor violation. EDMs are sensitive to new sources of CP violation in superpartner interactions, while flavor observables are sensitive to misalignment between fermion and scalar mass eigenstates and gain further sensitivity in the presence of additional CP violation.

There are two possible sources of EDMs in split supersymmetry. If the scalars are sufficiently light, there are one-loop diagrams contributing to EDMs involving loops of both scalar and fermionic superpartners, with CP violation arising through relative phases in SUSY-breaking soft parameters. If the new CP-violating phase is $\mathcal{O}(1)$, then the current electron EDM limit of $|d_e| < 8.7 \times 10^{-29}$ \cite{Baron:2013eja}  may be sensitive to scalars as heavy as $\tilde m \sim 2 \times 10^{5}$ GeV; the mass reach from limits on chromo-electric dipole moments is comparable.\footnote{Here we allow large flavor violation in the scalar sector to maximize the possible reach in $\tilde m$ \cite{McKeen:2013dma}; without large flavor violation the mass reach is an order of magnitude smaller.} Alternately, there may be contributions to EDMs from two-loop diagrams that only involve fermionic superpartners, which dominate the EDM signal if scalars are sufficiently heavy. Current EDM limits may be sensitive to fermionic superpartners as heavy as $10^3$ GeV. In both cases, this is the {\it maximum} expected mass reach, assuming maximal phases and radiative contributions. Thus split supersymmetry does not guarantee an EDM signal, since the mass range of scalars and fermions may lie beyond the sensitivity of EDM experiments, or the size of new CP violation may be too small. On the other hand, the observation of an anomalous EDM would place a suggestive upper bound on the mass scale of split supersymmetry.\footnote{Though even in this case, there is no guarantee that an anomalous EDM is an indication of supersymmetry; an anomalous EDM could be spoofed by various new degrees of freedom such as a CP-violating extended Higgs sector.}

The prospects for flavor violation are comparable to EDMs. The most sensitive observables involve CP violation in the kaon sector, where the maximal reach is currently $\tilde m \sim 10^6$ GeV assuming maximal CP and flavor violation in the down-type squark sector \cite{Bona:2007vi}. These bounds are unlikely to improve significantly in the near future, though limits on flavor violation in other meson sectors may eventually achieve comparable sensitivity. The strength of indirect probes such as EDMs and flavor violation is that observation of an anomalous signal would place an upper bound on the mass scale of split supersymmetry on the order of $\tilde m \lesssim 10^6$ GeV. This also highlights the primary weakness: there is a vast range of scales in split supersymmetry consistent with gauge coupling unification and the observed Higgs mass that lie beyond the reach of these precision observables.

\section{The Reach of Inflationary Observables}\label{sec:reach}

The Hubble scale during inflation, $H$, is the characteristic energy at which fields are excited from the vacuum.  Fields with mass $\lesssim H$ during inflation can contribute significantly to cosmological observables at later times.  If supersymmetry is relevant to inflation, then it is necessarily broken by the curvature of space-time, which is also set by $H$.  In this case we expect the masses of superpartners to be on the order of the Hubble scale during inflation but not parametrically larger. Concretely, in old minimal supergravity, the scalar partner of the inflaton receives a universal contribution of $m = 2 H$ \cite{Baumann:2011nm} from its curvature coupling alone, which is modified by model dependent contributions from gravity mediation.  As a result, superpartners will be produced from the vacuum for any value of $H$, provided that $H$ is the largest source of SUSY breaking.  This structure is further motivated by attempts to make inflation technically natural \cite{Baumann:2011nk}, although the large field range implied by the tensor amplitude \cite{Lyth:1996im, Baumann:2011ws} introduces additional challenges for producing a viable model of inflation.  

As we have seen, in the context of split supersymmetry the scale of SUSY breaking in the current vacuum is {\it at most} $\sqrt{F} \sim 10^{13-14}$ GeV and in general can be significantly smaller. For this reason, it is very plausible that supersymmetry is relevant to inflation, particularly if the Hubble scale during inflation is on the order of $H \sim 10^{14}$ GeV as suggested by BICEP2. Thus we expect to leverage the full power of supersymmetric signatures of inflation to probe scenarios of split supersymmetry. 

Given these general considerations, the presence of an extra scalar field $\sigma$ with a mass $m \sim H$ is a very natural consequence of SUSY in our universe.  However, in order to produce measurable deviations from the single-field consistency conditions, it must also couple to the inflaton, $\phi$, as was first studied in \cite{Chen:2009zp} under the name {\it quasi-single field inflation} (QSFI).  For our purposes it suffices to take the full Lagrangian for such a scalar field $\sigma$ to be 
\beq\label{equ:lagrangian}
{\cal L}_\sigma = -\tfrac{1}{2} [ \partial_\mu \sigma \partial^\mu \sigma + m^2 \sigma^2] - \mu \sigma^3 + \frac{\sigma}{\Lambda} [ \partial_\mu \phi \partial^\mu \phi - \langle \dot \phi \rangle^2] \ ,
\eeq
where we expect $m \sim H$ and otherwise remain agnostic about the size of $\mu$ and $\Lambda$. We have coupled $\sigma$ to $\phi$ derivatively in order to protect the approximate scale invariance of the observed power spectrum (which is enforced by an approximate shift symmetry, $\phi \to \phi+c$).  During inflation, $\dot \phi$ acquires a vev that introduces a tadpole for $\sigma$, which we have cancelled explicitly\footnote{In general, $\dot \phi$ is time dependent and therefore the above formula is not correct as written.  It is straightforward to enforce tadpole cancelation at all times by embedding this model in the effective field theory of inflation \cite{Cheung:2007st, Baumann:2011nk}.  See \cite{Assassi:2013gxa} for further discussion. } since we wish to study fluctuations around the minimum of the potential. Fluctuations in $\sigma$ are converted into fluctuations in $\phi$ through the $\Lambda$-suppressed coupling. Self-interactions of $\sigma$ therefore constitute the leading contribution to the bispectrum, with the shape of the non-gaussianity interpolating between local ($m \ll H$) and equilateral ($m \sim H$). Furthermore, the squeezed limit of the bispectrum bears the imprint of the nonzero $\sigma$ mass.

It is important to note that even Planck-suppressed interactions (i.e., $\Lambda \sim M_{\rm pl}$) are sufficient to generate a measurable signal \cite{Green:2013rd, Assassi:2013gxa}.  The requisite couplings were studied carefully in \cite{Assassi:2013gxa}, where for weak mixing ($\tfrac{\dot \phi}{ \Lambda} \ll H$) it was found that
\beq \label{eq:fnl}
\frac{\fNL^{\rm equil.}}{75} \sim  12 \, \frac{\mu}{H} \left(\frac{r}{0.2}\right)^{1/2}\left(\frac{\Mp}{\Lambda}\right)^3 \ .
\eeq
The existing constraint on $\fNL^{\rm equil.}$ from Planck is given by $\fNL^{\rm equil.} =-42 \pm 75$ (at 1$\sigma$), so we are already capable of measuring Planck-suppressed couplings for sufficiently large $\mu$, as was emphasized in \cite{Assassi:2013gxa}.  Large-scale structure surveys are expected to improve on these measurements through the galaxy bispectrum with potential sensitivity of $\Delta \fNL^{\rm equil.}\sim 10$ \cite{Sefusatti:2009xu} (or optimistically $\Delta \fNL^{\rm equil.} < 1$ \cite{Carrasco:2013mua}).

The natural question is then whether we generically expect additional scalars in split supersymmetry to possess the interaction terms in (\ref{equ:lagrangian}). Two very plausible, technically natural scenarios in which the above action can be generated are:
\begin{itemize}
\item If inflation is described by a single chiral superfield, $\sigma$ and $\phi$ can be the two real components of the complex scalar \cite{Baumann:2011nk}.  The mass $m\sim H$ for $\sigma$ is generated through gravity mediation and/or curvature couplings.  The self-interaction of $\sigma$ and the mixing between $\sigma$ and $\phi$ both arise through shift-symmetric irrelevant operators in the K\" ahler potential.\footnote{The three leading irrelevant operators of interest are $K \supset \frac{1}{\Lambda} (\Phi + \Phi^\dag)^3,  \frac{1}{\Lambda^3} X^\dag X (\Phi + \Phi^\dag)^3,$ and $\frac{1}{\Lambda^3} (\Phi + \Phi^\dag)^5$, where $\Phi$ is the superfield containing $\sigma, \phi$ while $X$ is a field seeding the value of the inflaton potential during inflation, $F_X \sim M_{\rm pl} H$ (see \cite{Baumann:2011nk} for further discussion). The first operator gives rise to the mixing term and also generates a self-interaction of $\sigma$ via curvature couplings. The second and third operator give rise to self-interactions for $\sigma$ directly. The contributions to $\mu/H$ from these operators are of order $H/\Lambda$, $M_{\rm pl}^2 H / \Lambda^3,$ and $\mathcal{O}(10^7) H^3 / \Lambda^3$, respectively, and all give rise to effects of similar numerical size. Note however that for the first operator, the same scale $\Lambda$ suppresses both the self-interaction and mixing terms; the naive $\Lambda$ 
required for an observable $f_{\rm NL}$ from this operator alone entails $\mathcal{O}(1)$ mixing. This lies outside the regime of validity of the weak mixing result (\ref{eq:fnl}) and a more detailed analysis of $f_{\rm NL}$ is required; see \cite{Assassi:2013gxa}. For the second and third operators, the $\Lambda$ can be slightly different from the scale suppressing mixing terms, generating an observable bispectrum while preserving the validity of the weak mixing result (\ref{eq:fnl}).} If these irrelevant operators are suppressed by powers of the same scale $\Lambda$ with comparable dimensionless coefficients, one typically expects $\Lambda \lesssim M_{\rm pl}/10$ is necessary for an appreciable signal. However, it is also possible for $\mu \sim H$ accidentally (i.e.~the irrelevant operator generating $\mu$ may have an anomalously large dimensionless coefficient), allowing an appreciable signal with $\Lambda \sim M_{\rm pl}$.

\item When there are multiple light chiral superfields, $\sigma$ and $\phi$ can arise from different superfields. In this case the potential for $\sigma$ is less constrained; both $\mu$ and $m$ can be generated via gravity mediation \cite{Assassi:2013gxa} and are naturally of the correct size to produce a measurable signal with $\Lambda \sim M_{\rm pl}$.
\end{itemize}
We will be agnostic about which scenario is more plausible. In the first case, the degrees of freedom and self-interactions are intrinsic to supersymmetric inflation with irrelevant operators in the K\" ahler potential, but the generic scale of irrelevant operators required for a signal is somewhat below $M_{\rm pl}$. In the second case, the degrees of freedom and self-interactions are not intrinsic to supersymmetric inflation (but are highly plausible ingredients), while the scale of irrelevant operators can naturally be $\mathcal{O}(M_{\rm pl})$.  Both are well motivated from different model building perspectives and lead to potentially observable signatures.

In either case, the signature of $m \sim H$ appears in the squeezed limit of the bispectrum.  The behavior in the squeezed limit was worked out analytically in \cite{Chen:2009zp}:
\beq\label{equ:qsfi}
\lim_{k_3 \to 0} B(k_1,k_2,k_3) = \frac{12}{5} \fNL^{\rm equil.} c_\alpha \times P_\zeta(k_1)P_{\zeta}(k_3) \left(\frac{k_3}{k_1} \right)^{\alpha}
\eeq
where $c_\alpha \sim {\cal O}(1)$. The observation of modest non-gaussianity and squeezed-limit scaling consistent with (\ref{equ:qsfi}) would then be suggestive evidence for supersymmetry at high scales.

In practice, we can use the QSFI template \cite{Chen:2009zp} given by
\beq\label{equ:template}
B^{\rm QSFI}(k_1,k_2,k_3) =  \frac{18}{5} \Delta_\zeta^4 \fNL^{\rm equil.} \frac{3^{3/2}}{N_{ -\alpha+3/2}[\frac{8}{27}]} \frac{N_{-\alpha+3/2}[\frac{8 k_1 k_2 k_3}{(k_1+k_2+k_3)^3}]}{[k_1 k_2 k_3(k_1+k_2+k_3)]^{\tfrac{3}{2}}}
\eeq
where $N_\nu[x]$ is the Neumann function.  It was shown in \cite{Sefusatti:2012ye} that the above ansatz is  in good agreement with the correct theoretical bispectrum in the equilateral, flattened and squeezed configurations away from $\alpha = 0$ and is therefore sufficient\footnote{In the limit $\alpha \to 0$, $\sigma$ becomes massless and the model becomes sensitive to physics at reheating.  Although the QSFI template does not agree with the analytic calculations in this limit, both are missing potential late time contributions which would contribute to the local shape.  Nevertheless, as our primary interest will be distinguishing $ \alpha \sim 1$ from $\alpha = 2$, we will not be concerned about inaccuracies around $\alpha = 0$.  
} for our purposes.  
\vskip 8pt
There is a strong analogy between the prospects for an inflationary signal of supersymmetry and low-energy probes such as EDMs and flavor violation. In the case of an inflationary signal, the degrees of freedom are intrinsic -- at the very least, supersymmetry demands a real scalar partner of the inflaton and protects the mass of other light scalars. Although the signal is not guaranteed to be accessible by next-generation experiments, there is nonetheless a wide range of well-motivated scenarios where the signal is appreciable. Non-observation of equilateral non-gaussianity and/or a squeezed limit with $\alpha < 2$ does not exclude supersymmetry, but observation of such a signal would provide a compelling indication for supersymmetry  at high scales. In the case of low-energy probes the degrees of freedom are also intrinsic -- supersymmetry demands superpartners of Standard Model fields -- but the signal is also not guaranteed to be accessible; CP-violating phases or flavor violation might be too small to observe even if the mass scales are within reach. As such, non-observation of low-energy anomalies likewise does not exclude supersymmetry at high scales, while observation of a signal would be highly suggestive. 

The distinct advantage of an inflationary signal over low-energy probes is that it remains sensitive to all scales of SUSY breaking allowed by split supersymmetry, and hence can probe SUSY breaking well beyond the reach of low-energy probes. That said, the finite reach of low-energy probes offers a complementary advantage: an anomalous EDM or FCNC could then be used to set a non-trivial upper bound on the scale of Standard Model superpartners. 
 
\section{Forecasts for an Ideal Experiment}\label{sec:forecast}

The best limits on primordial non-gaussianity to date come from the CMB via the Planck satellite \cite{Ade:2013ydc}.  The bounds from the CMB could be further improved through a future polarization-sensitive satellite mission \cite{Baumann:2008aq}.  In addition, the coming generation of large-scale structure (LSS) surveys can plausibly reach $\Delta \fNL^{\rm equil.} \sim 10$ \cite{Sefusatti:2009xu} via measurements of the bispectrum of tracers of the LSS.  In what follows we will focus on LSS as the parameters of the surveys are known, but similar considerations will apply to a future CMB experiment.

Our interest here is detecting deviations from the single field consistency condition (\ref{equ:singlecc}) of the form (\ref{equ:multi}).  A clean detection of $\alpha < 2$ unambiguously requires additional fields (beyond the inflaton) and for $\alpha > 0$ suggestively points to SUSY.  Although the specific model written in (\ref{equ:lagrangian}) makes predictions beyond the scaling in the squeezed limit, such as the detailed shape in equilateral configurations, these predictions are not robust to the inclusion of additional massive fields or self-interactions of the inflaton.  For this reason, we are interested in isolating the the scaling behavior in the squeezed limit to determine $\alpha$.  

Unfortunately, isolating the squeezed limit means our constraints will be weaker than if we used all the information at our disposal to constrain this specific model.  Forecasts for the measurement of $\alpha$ using the full QSFI template were performed in \cite{Sefusatti:2012ye, Norena:2012yi} and they indeed find stronger limits for a given $\fNLb$.  We will not follow the same strategy here to avoid using equilateral configurations to determine $\alpha$, as we wish to remain agnostic as to the underlying model of inflation.  In addition, the confidence at which one can rule out $\alpha =2$ is the most unambiguous signal of SUSY, but the point $\alpha =2$ is not a well-defined value for the QSFI template. For this reason, \cite{Sefusatti:2012ye, Norena:2012yi} cannot\footnote{The analysis of \cite{Norena:2012yi} also included a discussion of general scaling in the squeezed limit.  This analysis differs from ours in several respects and is therefore not directly comparable.} extend their forecasts beyond $\alpha = \tfrac{3}{2}$.  Although both types of analyses would be important to perform on real data, we believe our more conservative analysis would be required to definitely rule out single field inflation.

For the purpose of understanding our reach in $\alpha$, we will consider an ideal 3d measurement of the primordial correlation functions.  For cosmic variance limited measurements in the linear regime, this should be a good approximation for the experimental sensitivity.  Pushing $\Delta \fNL^{\rm equil.} < 10$ will ultimately require modeling (mildly) non-linear structure formation and it is less clear that these estimates will translate directly.  Nevertheless, these idealized estimates should provide a lower limit on the sensitivity of a real experiment.

Suppose we are given a fiducial model with some fiducial value of $\fNLb$ and $\bar\alpha$.  The likelihood function for $\fNL$ and $\alpha$ for these fiducial values, assuming scale invariance,  is given by \cite{Scoccimarro:2003wn, Babich:2004gb}
\beq\label{equ:lfunc}
-2 \log {\cal L} = \frac{V}{\Delta_\zeta^{6}} \int \frac{d^3 k_1}{(2\pi)^3} \frac{1}{(4\pi^2)} \int_{1/2}^{1-\epsilon_*} d x_2 \int_{1-x_2}^{x_2} d x_3\, ( B(1,x_2,x_3) - \bar B(1,x_2,x_3) )^2  x_2^4 x_3^4
\eeq
where $x_{2,3} = \frac{k_{2,3}}{k_1}$, $V = (2\pi)^3 / k_{\rm min}^3$ is the spatial volume of the survey and $\bar B$ and $B$ are the bispectra with the fiducial and measured values of parameters, respectively.  The parameter $\epsilon_*$ denotes the most squeezed configurations available in the survey, $\epsilon_* \equiv \frac{k_{\rm min}}{k_{\rm max}}$.  

In this parameterization, the squeezed limit corresponds to $x_2 \to 1$ and $x_3 \to 0$.  It is easy to see that for $\bar\alpha > 0$, the dominant contribution to the likelihood function is not from the squeezed limit but from more equilateral configurations ($x_2 \sim x_3 \sim 1$).  In practice, this means this first signature of SUSY would be the detection of $\fNL^{\rm equil.}$.  In addition, for $\alpha > 0$ we are mildly sensitive to extremely squeezed configurations, which tend to differ significantly depending on the type of observation.  As result, our forecasts should be fairly robust away from $\alpha= 0$.  

Our goal here is to understand the limits that we can place on $\alpha$ without reference to the equilateral configuration.  If we isolate only the squeezed configurations with momenta in the range $\epsilon_* < x_3 < \epsilon$, from our ansatz in  (\ref{equ:qsfi}), we have
\beq
-2 \log {\cal L} =\frac{ (\fNL-\fNLb)^2}{\sigma_{\fNL}^2} + \frac{1}{\sigma_{\fNL}^2}  \int _{1-\epsilon}^{1-\epsilon_\star} \frac{d x_2}{x_2^2} \int_{1-x_2}^{\epsilon}  \frac{d x_3}{x_3^2} \, \left[ \fNL C_\alpha \left( \frac{x_3}{x_2} \right)^{\alpha} -\fNLb C_{\bar \alpha} \left( \frac{x_3}{x_2} \right)^{\bar\alpha} \, \right]^2 \,
\eeq
where $\sigma_{\rm \fNL}$ is the 1-$\sigma$ error placed on $\fNL^{\rm equil.}$ from equilateral measurements and $C_\alpha \equiv c_\alpha \fNL / \sqrt{B \cdot B}$ with $B\cdot B \equiv \int_{1/2}^{1-\epsilon} dx_2 \int^{x_2}_{\epsilon} dx_3 B(1,x_2,x_3)^2 x_2^4 x_3^4$.  

\begin{figure}[t]
\begin{center}
\includegraphics[width=.6 \textwidth]{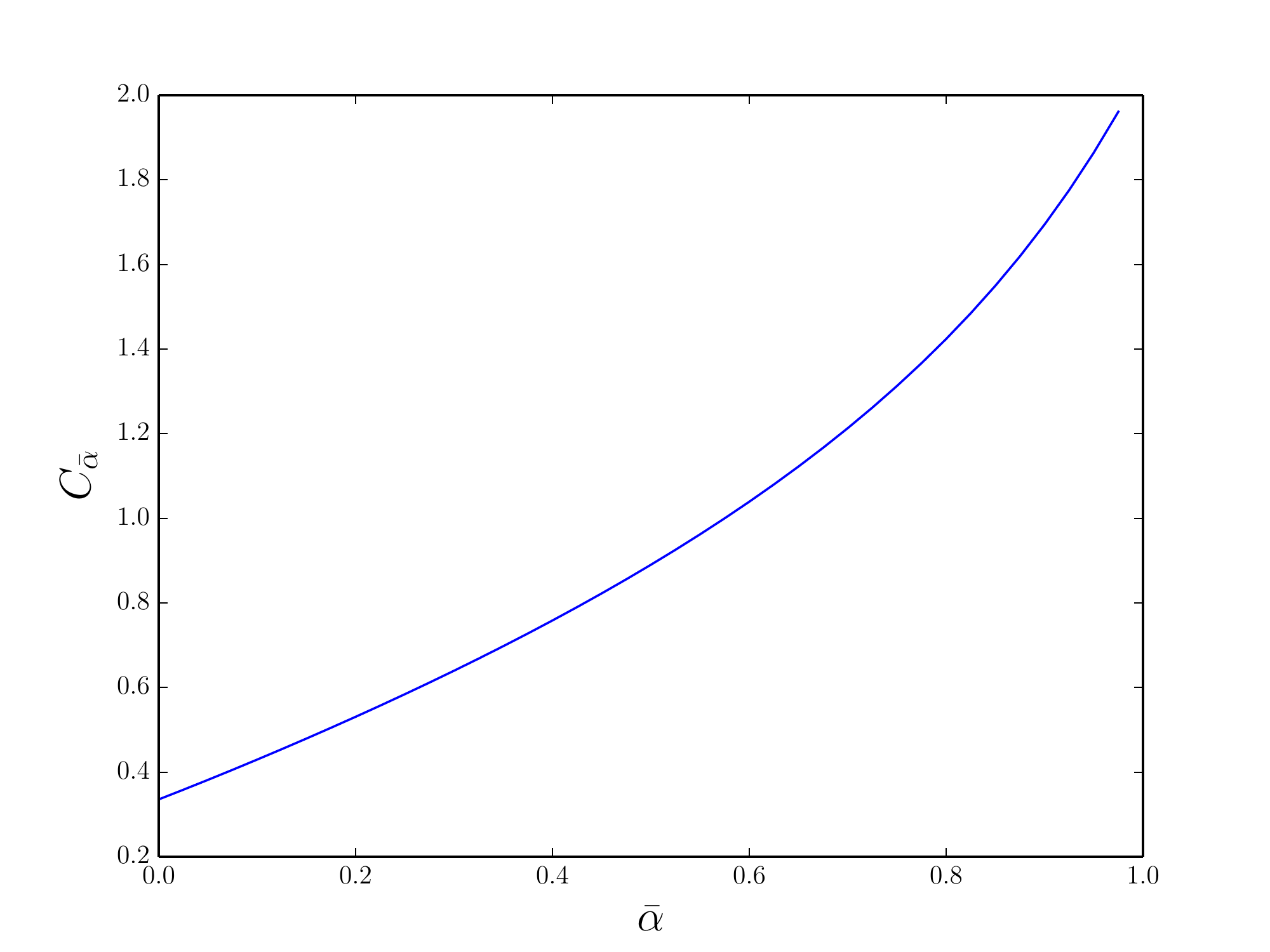}
\caption{ \label{fig:norm} \small\it  The relative normalization of the amplitude of the likelihood function in the squeezed and equilateral limits, as defined by $C_{\bar\alpha}$.  The function was computed for varying fiducial values of $\bar \alpha$ using the QSFI template in (\ref{equ:template}). }
\end{center}
\end{figure}

It is crucial that our determination of $\alpha$ comes only from the exponent in the squeezed limit.  Therefore, to determine sensitivity to $\alpha$ we should marginalize over $\fNL$ and $C_\alpha$.  We will take a flat prior on $\fNL$ for simplicity.  This choice is reasonable because the likelihood function will be dominated by the leading term under the assumption that we detect the equilateral shape.  We will also take a flat prior on $\tilde C_\alpha \equiv \fNL C_\alpha$ in order to remain agnostic as to the relationship between $\fNL^{\rm equil.}$ and the amplitude in the squeezed limit.  We will assume the fiducial $C_{\bar \alpha}$ is the one from QSFI, computed using the QSFI template (\ref{equ:template}) and shown as a function of $\bar \alpha$ in figure \ref{fig:norm}.

\begin{figure}[t]
\begin{center}
\includegraphics[width=.7 \textwidth]{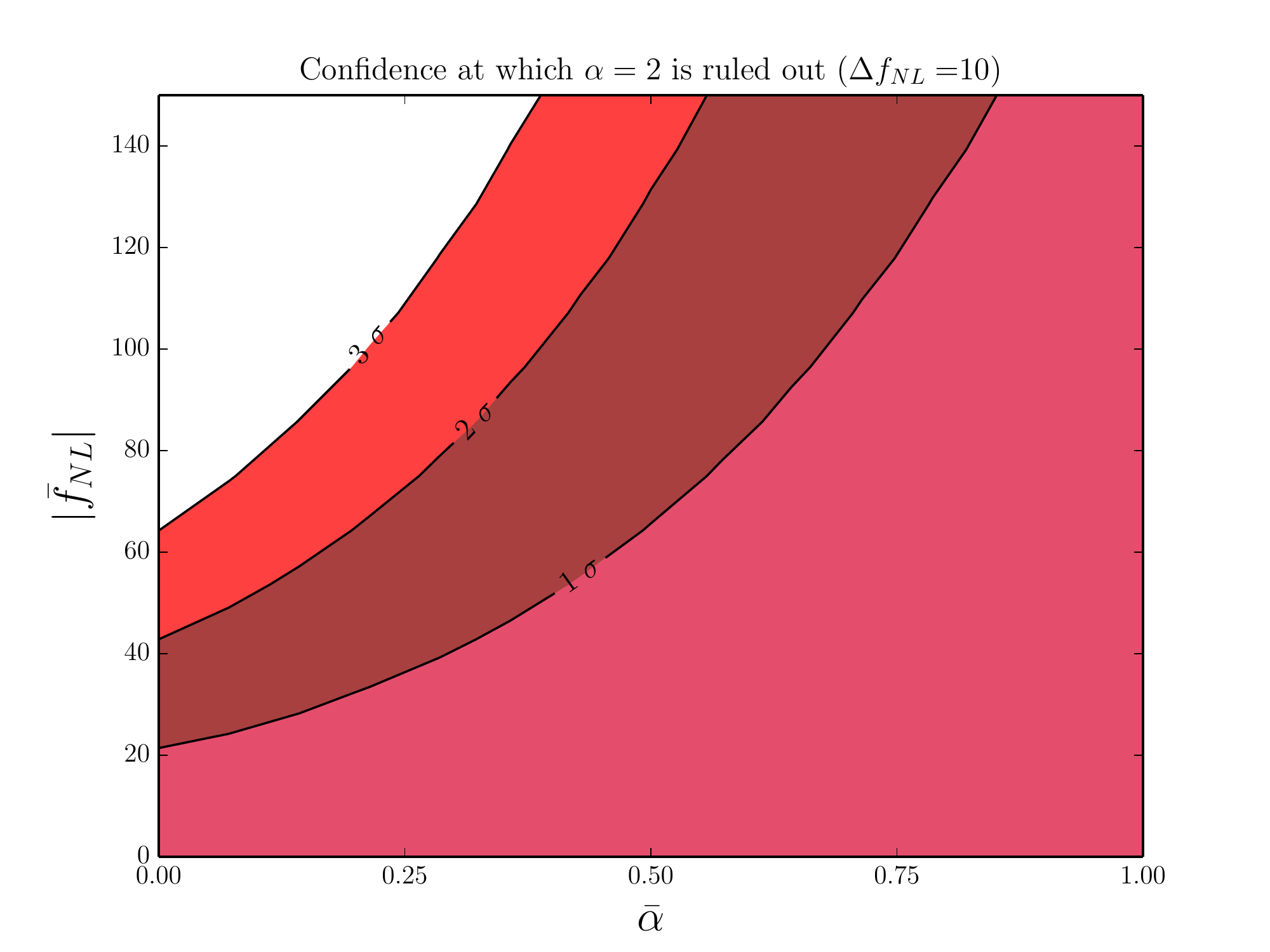}
\caption{ \label{fig:contours1} \small\it  The confidence level at which $\alpha = 2$ can be ruled out as a function of varying $\bar f_{NL}$ and $\bar \alpha$. The forecasted contours use $\sigma_{\fNL} = 10$, $\epsilon = 10^{-1}$ and $\epsilon_* = 3\times10^{-3}$.  }
\end{center}
\end{figure}

\begin{figure}[t]
\begin{center}
\includegraphics[width=.7 \textwidth]{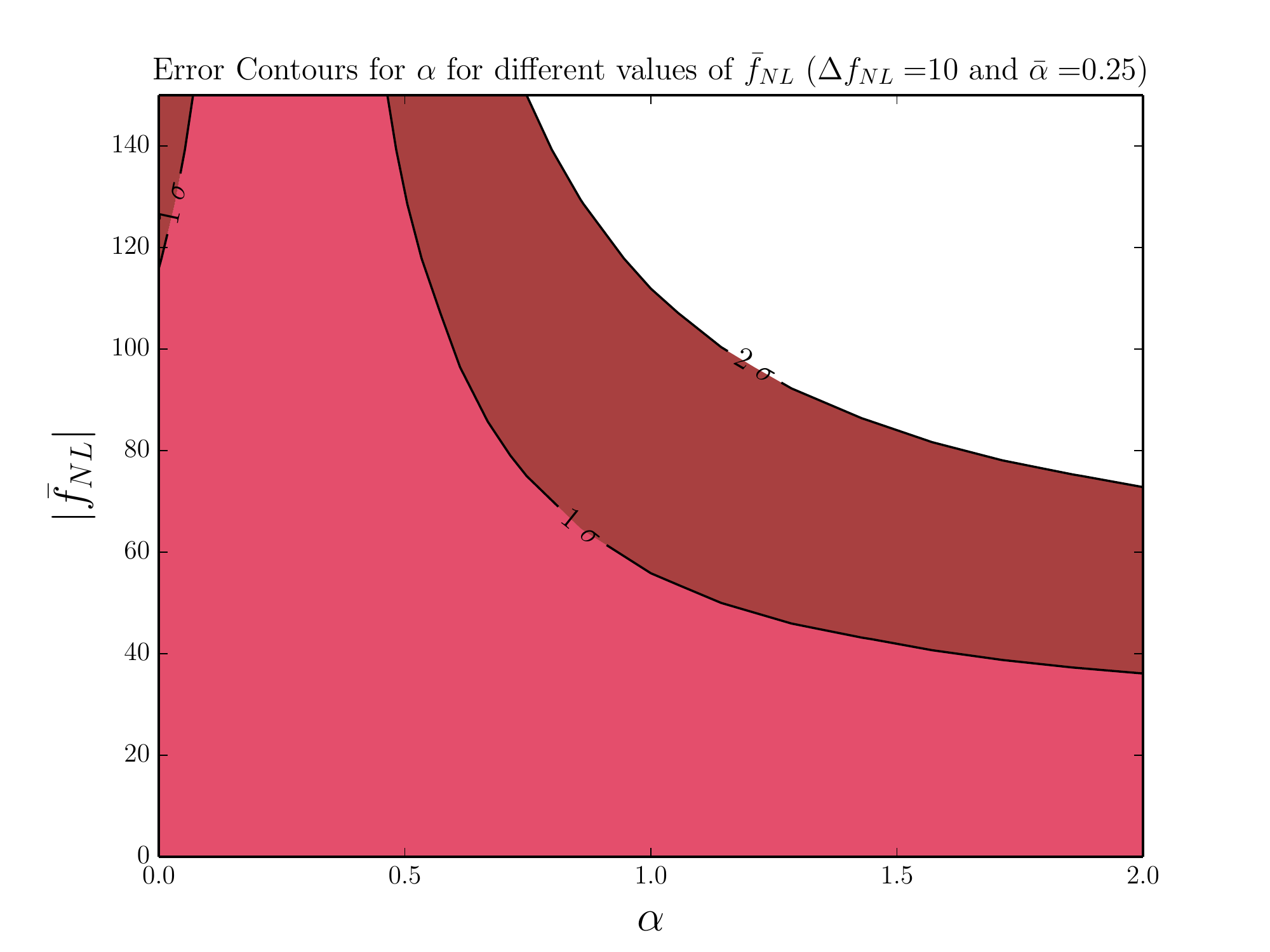}
\caption{ \label{fig:contours2} \small\it  The 1$\sigma$ and 2$\sigma$ error contours on the measurement of $\bar \alpha = 0.25$ for varying values of $\bar f_{NL}$. The forecasted contours use $\sigma_{\fNL} = 10$, $\epsilon = 10^{-1}$ and $\epsilon_* = 3\times10^{-3}$.   While the plot shows clear measurements of $\alpha < 2$, it appears to show results compatible with $\alpha = 0$.  This is an artifact of our flat prior on $\tilde C$ and our assumption that the dominant signal is in the equilateral template.  For $\bar\alpha \sim 0$, the squeezed limit dominates the signal to noise and we would see this as a detection of $\fNL^{\rm local}$. }
\end{center}
\end{figure}

The projected reach in $\alpha$ is shown in figures \ref{fig:contours1} and \ref{fig:contours2} for $\sigma_{\fNL} = 10$.  We assume the squeezed limit applies for $\epsilon = 10^{-1}$ and we take $\epsilon_* = \frac{k_{\rm min}}{k_{\rm max}} \sim 3 \times 10^{-3}$.  The choice of $\epsilon_*$ is based on expectations from the Euclid survey \cite{Laureijs:2011gra} volume of $V = 108 \, h^{-3} \, {\rm Gpc}^3$ and $k_{\rm max} \sim 0.4 \, h \, {\rm Mpc}^{-1}$ (which is fairly conservative for $1<z < 2$).  Taking advantage of scale-dependent bias \cite{Dalal:2007cu, Schmidt:2010gw} in the bispectrum may ultimately improve this effective range \cite{Baldauf:2010vn}, but we will take this more conservative choice.  Given that the signal is not concentrated in the squeezed limit, bias should not play a huge role beyond determining $\sigma_{\fNL}$.

In figure \ref{fig:contours1}, we address the key question of how well $\alpha = 2$ can be excluded as a function of the fiducial values $\bar f_{NL}$ and $\bar \alpha$.  For values of $\bar f_{NL}$ consistent with current Planck limits $|\fNL| < 117$ (1$\sigma$), $\alpha = 2$ can be ruled out by as much as $3 \sigma$ for small values of $\bar \alpha$.  Furthermore for $|\fNL| >70$, discrimination is possible out to $\bar \alpha = 0.25$ with more than 2$\sigma$, which corresponds to a scalar of mass 
\beq
m \lesssim 0.8 H \ ,
\eeq
validating the proposal that a scalar of mass $m \sim H$ can be probed by cosmological observables within the next generation of LSS and CMB polarization surveys. In figure \ref{fig:contours2}, we then show the expected precision with which the borderline case of a fiducial value $\bar \alpha = 0.25$ can be measured as a function of $\bar f_{NL}$.  

\begin{figure}[t]
\begin{center}
\includegraphics[width=.7 \textwidth]{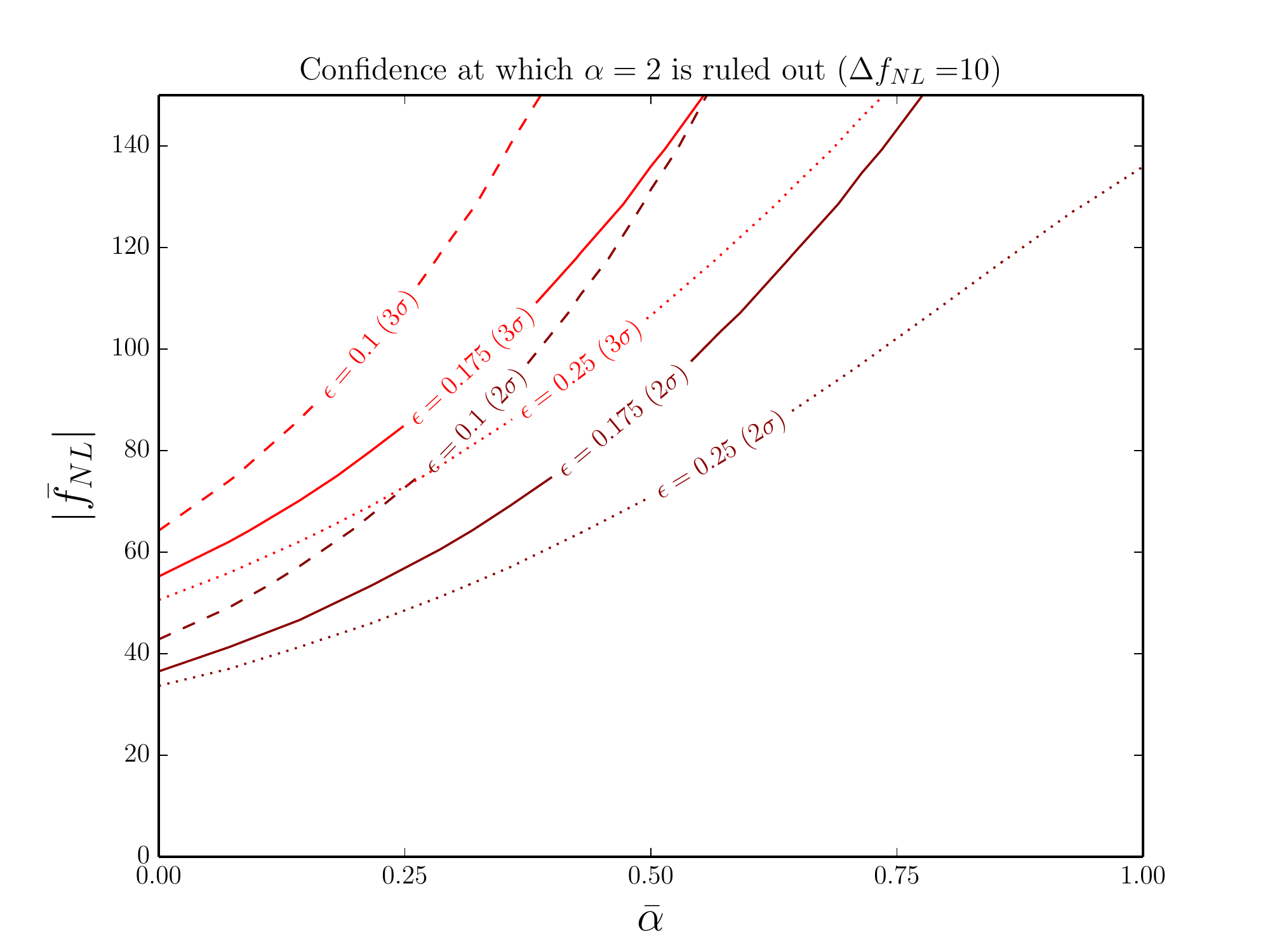}
\caption{ \label{fig:eps} \small\it  The confidence level at which $\alpha = 2$ can be ruled out as a function of varying $\epsilon$, $\bar f_{NL}$ and $\bar \alpha$. The forecasted contours use $\sigma_{\fNL} = 10$ and $\epsilon_* = 3\times10^{-3}$.  The 2$\sigma$ and 3$\sigma$ contours are shown for $\epsilon = 0.1,0.175,0.25$.  Our ability to rule out $\alpha =2$ improves significantly as we include more information from equilateral configurations.}
\end{center}
\end{figure}

It bears emphasizing that our forecasts are relatively conservative.   By ignoring equilateral configurations in measurements of $\alpha$, we are intentionally neglecting a lot of information.  For example, our choice of $\epsilon \equiv\left( \frac{k_1}{k_3}\right)_{\rm min}$ defining the minimum required squeezing to be included in the measurement of $\alpha$ is somewhat arbitrary.  In figure \ref{fig:eps} we show how our ability to rule out $\alpha =2$ is sensitive to our choice of $\epsilon$.  As we increase $\epsilon$, our ability to rule out $\alpha=2$ improves rapidly because the signal to noise is concentrated in the equilateral configurations.  Even a modest change in $\epsilon$ substantially improves our discrimination.

Despite our somewhat conservative choice of $\epsilon = 0.1$, the galaxy bispectrum of a single survey (e.g. Euclid or BOSS) is capable of a detection of equilateral non-gaussianity ($>5\sigma$) and ruling out $\alpha =2$ at 3$\sigma$ for values of $\fNL^{\rm equil.}$ consistent with Planck at 1$\sigma$.  Given such a detection, one could gain more significance for the detection of $\alpha < 2$ through combined analysis with other probes (e.g. scale dependent bias, multiple tracers, CMB) and improved modeling of mildly nonlinear scales.

\section{What Can Fake the Signal?} \label{sec:fakes}

Although the bispectrum is sensitive to SUSY during inflation, one might also wonder to what degree this is a unique signature.  In other words, if we were to make a detection of $\alpha < 2$, how confident would we be that there is supersymmetry in the universe?

Clearly, fully supersymmetrizing the Lagrangian in (\ref{equ:lagrangian}) entails adding fermions to complete the appropriate supermultiplets.  While the fermions will also contribute to observables, their contributions are typically suppressed by a power of $\Delta_\zeta$.  As a result, given current bounds on non-gaussianity, future surveys are not sufficiently sensitive to detect fermions with couplings related by SUSY.  Rather, the crucial role of SUSY in our discussion was to explain the origin of a scalar with $m \sim H$ naturally.  If one is simply willing to fine-tune a scalar to have the same Lagrangian and $m \sim H$ without radiative protection from a symmetry, then the dominant signal will be the same, although it is difficult to imagine even an anthropic reason for such a tuning (notice that $\sigma$ cannot be the Higgs because the linear coupling is not compatible with gauge invariance).  Some supersymmetric scenarios are even harder to spoof. For example, in the scenario with multiple light chiral multiplets, we expect many scalars to have $m \sim H$, not simply one. If more than one scalar couples to the inflaton, the measured bispectrum will differ from that of a single scalar. Without supersymmetry, the existence of many such light\footnote{Massless scalars can be protected by a shift symmetry but produce the local shape ($\alpha = 0$).} scalars seems highly implausible .

The signatures of models with many fields with $m \sim H$ will ultimately depend on the spectrum of masses and couplings.  The dominant contribution to the equilateral configurations will arise from the fields with the largest couplings to $\phi$ (and to a lesser extent, size of $\mu$).  On the other hand, the signal to noise in the squeezed limit will be power law suppressed for the more massive fields.  The dominant contribution in both equilateral and squeezed configurations may therefore arise from different fields entirely.  For some distributions of couplings, it is possible that equilateral configurations will be enhanced by the number of fields, potentially making the squeezed limit of the lightest field unmeasurable, but all such conclusions are model dependent.  

The signature in the squeezed limit can also be mimicked by anomalous dimensions, as was shown in \cite{Green:2013rd}.  In this case, rather than a massive scalar coupled to $\phi$, we have some operator ${\cal O}$ with a scaling dimension $\Delta$.  In the presence of ${\cal O}$ one finds the same behavior in the squeezed limit with $\alpha = \Delta$ when $\Delta \leq 2$.

\begin{figure}[t]
\begin{center}
\includegraphics[width=.45\textwidth]{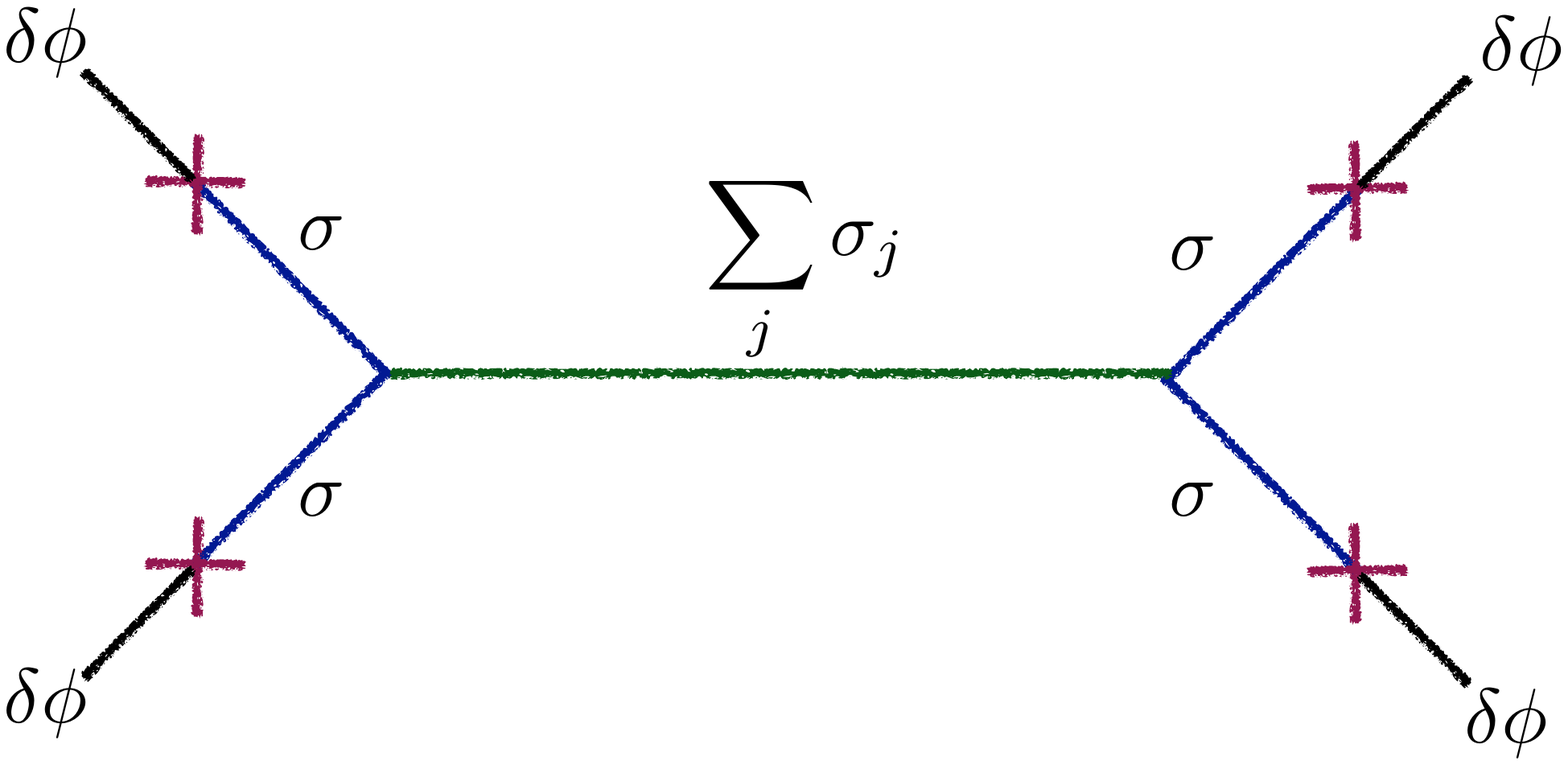}
\includegraphics[width=.45 \textwidth]{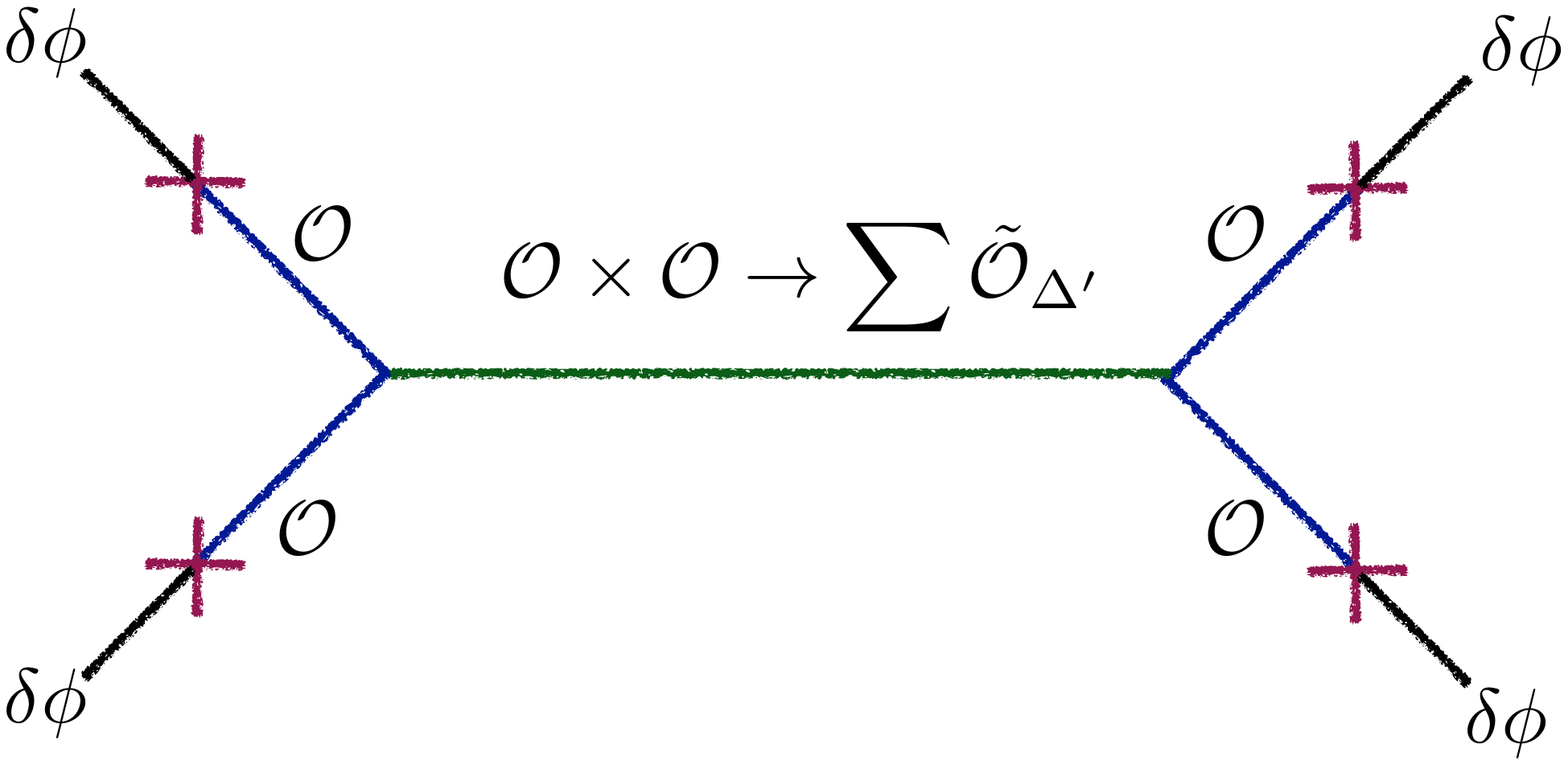}
\caption{ \label{fig:diagrams} \small\it  Schematic Feynman diagrams for the trispectrum of SUSY (left) and anomalous dimensions (right).  The SUSY trispectrum gets contributions from the exchange of superpartners with $m_i \lesssim H$ where $\sigma$ is the linear combination that couples to $\phi$ and $\sigma_i$ are all superpartners that are coupled via cubic interactions with $\sigma$ (including $\sigma$ itself).  When we have some operator ${\cal O}$, the trispectrum gets contributions from every operator $\tilde {\cal O}_{\Delta'}$ that appears in the operator product expansion of ${\cal O}$ with itself.  For interacting theories, these contributions should differ significantly from the weakly coupled scalars. }
\end{center}
\end{figure}

In principle, the trispectrum offers the opportunity to distinguish these models.  The trispectrum always gets contributions from exchanging the inflaton between two bispectra.  However, when there are extra fields involved, there are additional contributions to the trispectrum from the exchange of these fields \cite{Assassi:2012zq}, as shown in figure \ref{fig:diagrams}.  For SUSY, these contributions come from the exchange of $\sigma$ or any other superpartner.  In this case, the trispectrum signal-to-noise is suppressed only by $\frac{\mu}{H}$ relative to the bispectrum and is potentially detectable.  A similar phenomenon occurs when we couple to an operator ${\cal O}$, but now every operator in the operator product expansion contributes to the trispectrum \cite{Green:2013rd}.  These contributions are generically quite different from those of a weakly coupled scalar and could be used to distinguish between the two scenarios.

A final issue is the possibility that inflation took place in an excited state (i.e.~not in the Bunch-Davies vacuum).  As discussed in, e.g. \cite{Holman:2007na,Ganc:2011dy,Chialva:2011hc}, excited states lead to violations of the single field consistency condition and can produce measurable signals in the squeezed limit.  However, achieving more than two orders of magnitude of scale-invariant perturbations is challenging due to back-reaction constraints \cite{Flauger:2013hra}.  The models can often be distinguished in the CMB which covers three orders of magnitude in scale.  

As a whole, the prospects are closely analogous to low-energy probes such as EDMs and flavor violation. A signal in low-energy channels could also be spoofed by physics unrelated to SUSY, such as an extended Higgs sector, but the scales and interactions relevant for the signal are highly suggestive of supersymmetry and further measurements may disentangle various alternatives.

\section{Outlook} \label{sec:outlook}

The observed Higgs mass and apparent success of supersymmetric gauge coupling unification suggests that supersymmetry is relevant during the inflationary era, even if it is broken well above the weak scale.  In this work we have demonstrated the potential for next-generation measurements of cosmological observables to probe supersymmetry at high scales. A variety of generic and highly-motivated supersymmetric scenarios give rise to the signals of quasi-single field inflation -- namely, equilateral non-gaussianity with a non-trivial squeezed limit bearing the imprint of an additional scalar with $m \sim H$ during inflation. The existence of such an additional scalar can be distinguished from single-field scenarios in next-generation measurements. Observation of a signal would be suggestive of supersymmetry at high scales, in close analogy with the prospects of low-energy probes such as EDMs and flavor violation.

The key observable in the search for SUSY is a bispectrum with an equilateral shape.  This analysis is performed optimally in the CMB \cite{Ade:2013ydc} but a similar analysis will be required in LSS surveys (which has not been performed to date).  Galaxy surveys present a number of complications beyond the CMB and their ultimate reach remains to be determined, somewhat analogous to the challenges presented by the LHC compared to LEP.  The amount of information in LSS vastly exceeds the CMB, provided we can address these issues.

Although the recent BICEP2 measurement of tensor modes provides strong motivation for probing supersymmetry during inflation, our analysis remains relevant even if the central value of $r$ changes substantially in future measurements.  If the central value of $r$ is lowered, the scale $\Lambda$ probed by LSS and CMB polarization measurements changes according to (\ref{eq:fnl}). Our forecasting for $\alpha$ is unchanged. Should the signal of interest be observed -- namely, modest non-gaussianity with evidence in the squeezed limit for a scalar of mass $m \sim H$ -- the interpretation in terms of split supersymmetry remains valid. Moreover, now an upper bound could be set on the scale of supersymmetry breaking in the current vacuum, much in the same way that observation of an anomalous EDM could place an upper bound on the scale of soft masses.\\

\noindent {\bf Acknowledgments}\\
We are grateful to Nima Arkani-Hamed, Valentin Assassi, Daniel Baumann, and Liam McAllister for useful conversations. NC is supported by the DOE under grants DOE-SC0010008, DOE-ARRA-SC0003883, and DOE-DE-SC0007897.  The research of D.G.~is supported in part by the Stanford Institute for Theoretical Physics and by the U.S. Department of Energy contract to SLAC no.\ DE-AC02-76SF00515. 

\newpage
 \begingroup\raggedright

\end{document}